\documentclass[12pt]{article}
\usepackage{graphicx}

\newcommand\pubdate{\today}

\def\roch{Department of Physics $\&$ Astronomy\\
University of Rochester,
 Rochester, NY 14627-0171}
\def\support{\footnote{Work supported by the U.S. National Science Foundation.}}

\def\Title#1{\begin{center} {\Large #1 } \end{center}}
\def\Author#1{\begin{center}{ \sc #1} \end{center}}
\def\Address#1{\begin{center}{ \it #1} \end{center}}

\newcommand\pubblock{\rightline{\begin{tabular}{l} \\
         \pubdate  \end{tabular}}}
\newenvironment{Abstract}{\begin{quotation}  }{\end{quotation}}
\newenvironment{Presented}{\begin{quotation} \begin{center} 
             PRESENTED AT\end{center}\bigskip 
      \begin{center}\begin{large}}{\end{large}\end{center} \end{quotation}}
\def\Acknowledgements{\bigskip  \bigskip \begin{center} \begin{large}
             \bf ACKNOWLEDGEMENTS \end{large}\end{center}}
\usepackage{relsize}
\def\babar{\mbox{\slshape B\kern-0.1em{\smaller A}\kern-0.1em 
  B\kern-0.1em{\smaller A\kern-0.2em R}}}
\def \br{{\cal B}}

\def\beq{\begin{equation}}
\def\eeq#1{\label{#1}\end{equation}}
\def\eeqn{\end{equation}}

\def\beqa{\begin{eqnarray}}
\def\eeqa#1{\label{#1}\end{eqnarray}}
\def\eeqan{\end{eqnarray}}

\let\bar=\overbar

\def\Dslash{\not{\hbox{\kern-4pt $D$}}}
\def\dslash{\not{\hbox{\kern-2pt $\del$}}}

\def\msb{{\bar{\ssstyle M \kern -1pt S}}}

\begin{document}
\begin{titlepage}
\pubblock

\vfill
\Title{Search for $D^0\to\gamma\gamma$ at BESIII}
\vfill
\Author{ Hajime Muramatsu\support}
\Address{\roch}
\vfill
\begin{Abstract}
Based on the ${\sim}2.9$ fb$^{-1}$ data set taken at the nominal mass of
$\psi(3770)$, we report our preliminary result on a search for $D^0\to\gamma\gamma$.
We find no significant signal and set an upper limit on
$\br(D^0\to\gamma\gamma)/\br(D^0\to\pi^0\pi^0)<5.8\times10^{-3}$ at
$90\%$ confidence level.
\end{Abstract}
\vfill
\begin{Presented}
The $5^{th}$ International Workshop on Charm Physics \\
Honolulu, Hawai'i, May 14--17, 2012
\end{Presented}
\vfill
\end{titlepage}
\def\thefootnote{\fnsymbol{footnote}}
\setcounter{footnote}{0}
%
\section{Why $D^0\to\gamma\gamma$ ?}

In the Standard Model (SM), flavor-changing neutral currents (FCNC)
occur only at the loop level, where they are highly suppressed by the
GIM mechanism~\cite{gim1970}. The decay of $D^0 \rightarrow \gamma
\gamma$ must be produced by charm-changing neutral currents.
From the short distance contributions, the decay rate for $D^0\rightarrow \gamma
\gamma$ is predicted to be $3\times
10^{-11}$~\cite{greub1996,fajfer2001,burdman2002}. However, the long
distance contributions significantly enhance the decay rate which is
estimated to be $(1-3)\times 10^{-8}$~\cite{fajfer2001,burdman2002}.
This FCNC decay could be enhanced by new
physics (NP) effects which lead to contributions at loop
level~\cite{Prelovsek2001, bigi2010}.
For instance, in the framework of the Minimal Supersymmetric Standard Model
(MSSM), the calculation shows that the decay rate for $c\rightarrow
u \gamma$ transition could be $6\times 10^{-6}$, 
which is one to two orders of magnitudes enhanced relative to the SM rate, by considering
gluino exchange~\cite{Prelovsek2001}. 

Experimental searches for $D^0\rightarrow \gamma \gamma$
were performed by the CLEO~\cite{cleo} and $\babar$~\cite{babar} experiments based on data
samples collected at the $\Upsilon(4S)$ peak. They found
no significant signals. 
The latter experiment yields the most stringent experimental upper
limit to date on the 
$\br(D^0\rightarrow \gamma\gamma)$, $2.4\times 10^{-6}$ at 90\% confidence level (C.L.).

Here I report a preliminary result of our search for
$D^0 \rightarrow \gamma\gamma$ based on a sample acquired at the
BEPCII energy-symmetric $e^+e^-$ collider with the BESIII detector~\cite{besiii}.
This sample was collected at $\sqrt{s}=3.773$~GeV with an
integrated luminosity of ${\sim}2.9$ fb$^{-1}$ in which the background level
at this open charm threshold is expected to be substantially lower than that at the $\Upsilon(4S)$ peak.


\section{Analyses}

One of the major backgrounds in the analysis of $D^0\to\gamma\gamma$
comes from
$D^0\to\pi^0\pi^0$. I first describe our study on the decay of
$D^0\to\pi^0\pi^0$, then move on to discuss our search for
$D^0\to\gamma\gamma$.
In the end, we give our result in terms of
a ratio of branching fractions,
$\br(D^0\to\gamma\gamma)/\br(D^0\to\pi^0\pi^0)$.


\subsection{$D^0\to\pi^0\pi^0$}

Each $\pi^0$ candidate is constructed from a pair of {\it good
  photons}, each of which is defined as:
it is based on a shower in the electromagnetic CsI crystal
    calorimeter (EMC) of the BESIII detector~\cite{besiii};
should not match with any charged tracks
    reconstructed by the main drift chamber;
must be reconstructed within
         barrel ($|\cos{\theta}|<0.80$) or  endcap
         ($0.86<|\cos{\theta}|<0.93$) sections of the EMC
         where $\theta$ is a polar angle with respect to the positron
         beam direction;
deposited energy of the shower must be at least
$25$ $(50)$~MeV when it is reconstructed in the barrel (endcap) section. 
Once a pair is found to be a $\pi^0$ candidate
($110<M_{\gamma\gamma}<150$~MeV$/c^2$), it is kinematically constrained to the
known mass~\cite{pdg} from which the resultant constrained momentum of $\pi^0$
is used for the rest of the analysis.
These $\pi^0$ candidates are also tested for the possibility that one of
their showers could form a $\pi^0$ with some other shower, and are
rejected if any other pairing were more consistent with a $\pi^0$ mass.
Then, we form $\Delta E$ $(\equiv E_{\pi^0\pi^0} - E_{beam})$ for every two $\pi^0$
candidates and require $(-60<\Delta E<30)$~MeV.
In the end, we extract our signal from a distribution of
beam-constrained mass,
$M_{bc}$ $\equiv$ $\sqrt{E^2_{beam} - \mathbf{p_D}^2c^2}$ where $E_{beam}$
is the beam energy and $\mathbf{p_D}$ is the $D^0(\to\pi^0\pi^0)$
candidate momentum.

\begin{figure}[htb]
\centering
\includegraphics[height=4.0in,trim=0mm 0mm 0mm 33mm,clip]{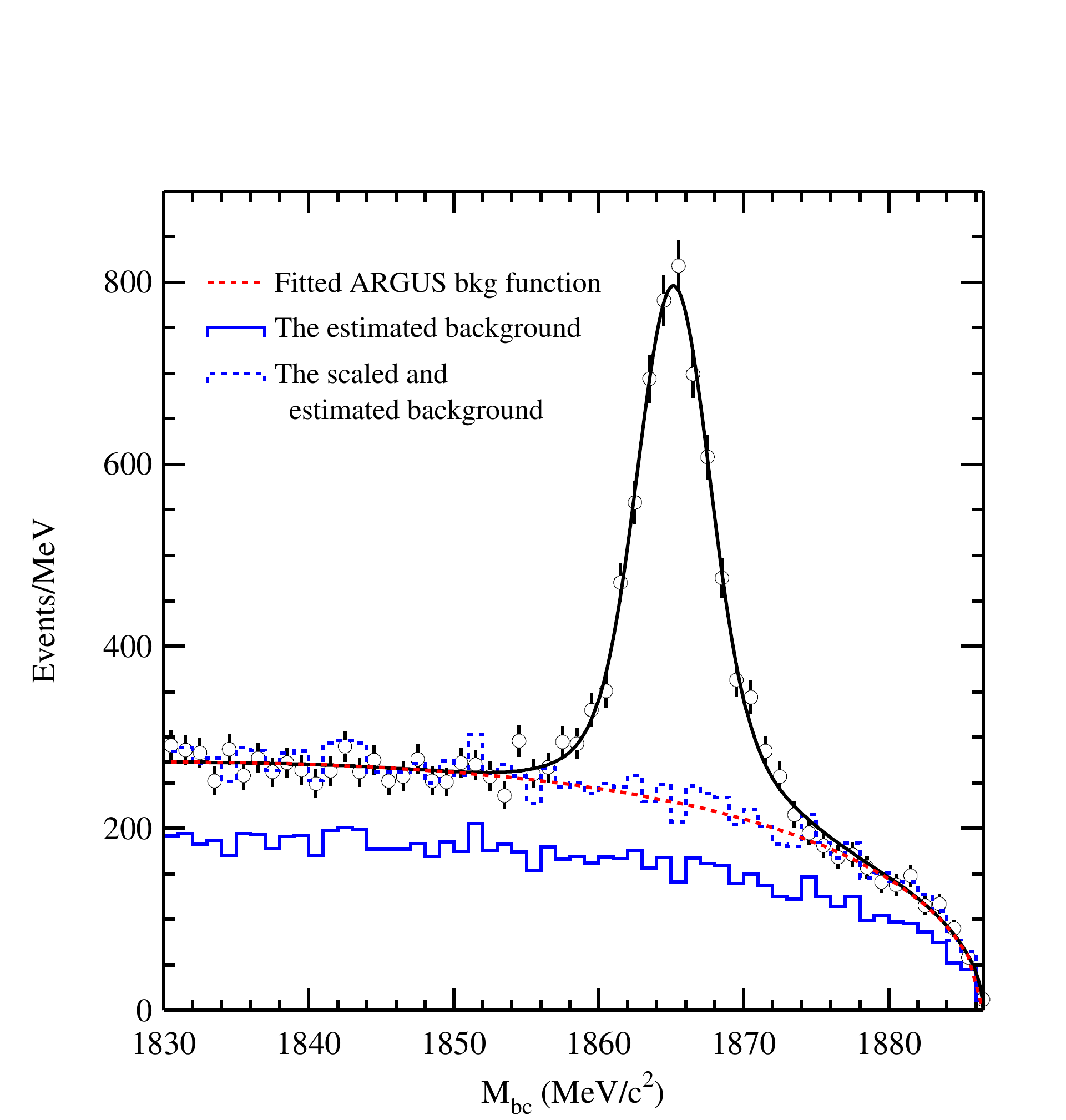}
\caption{A fit to the $D^0\to\pi^0\pi^0$ candidate $M_{bc}$ distribution based on
the $\psi(3770)$ data sample. The black points are data, the
black-smooth curve represents the overall fit (signal plus
background), and the red-dashed curve corresponds to the fitted
background shape. The blue-solid histogram represents the expected
background shape and size based on our MC samples
while the blue-dotted histogram is a fit to the data based on this expected
MC-based background shape.
}
\label{fig:pi0pi0_nominal_dt_fit}
\end{figure}

We perform a maximum likelihood fit to the resultant $M_{bc}$
distribution based on our data sample as shown in 
Figure~\ref{fig:pi0pi0_nominal_dt_fit}.
The signal shape is represented by a double Gaussian and its background is described by
an ARGUS background function. In this figure, we also overlay the
expected background component based on our Monte Carlo (MC) samples, represented by the
blue-solid histogram, which apparently fails to describe the observed
background level in data. We attribute this discrepancy to the
incomplete simulation of continuum process
under the $\psi(3770)$ peak.

The fit yields
$4081\pm117$ signal events with $52\sigma$
statistical significance. The significance is obtained by
$\sqrt{-2\times\ln{({\cal L}_{wo}/{\cal L}_w)}}$ where
${\cal L}_{wo}$ and ${\cal L}_w$ are the likelihood values obtained from a fit
without and with the signal shapes.
The $\chi^2$ for the fit is 42.0 for 56 data points
(minus the 5 floated parameters).

We estimate possible systematic uncertainties on the extracted yields that include
fit range, background shape, the requirement on $\Delta E$, and
reconstruction of $\pi^0$.
The single largest source of systematic uncertainty is due to the
assumed signal shape. While we use a double Gaussian to represent the
signal shape, the shape of a double Gaussian was fixed based on our
MC simulation which cannot describe the data well.
This is currently under the investigation as of this writing.
Table~\ref{tab:pi0pi0syst} shows a summary of dominant sources of
the estimated systematic uncertainties.

\begin{table}[htbp]
\caption{Systematic uncertainties on the extracted yields
of $D^0\to\pi^0\pi^0$.}
\label{tab:pi0pi0syst}
\begin{center}
\scalebox{1.0}
{
\begin{tabular}{c| c}
Sources & Rel. error ($\%$) \\
\hline
Signal shape & $\pm8.7$ \\
$\pi^0$ recon.    & $\pm2.0$ \\
MC stat.          & $\pm0.3$ \\
\hline
Total             & $\pm8.9$ \\
\hline
\end{tabular}
}
\end{center}
\end{table}

\noindent With the total reconstruction efficiency of
$(23.3\pm0.1)\%$, the preliminary efficiency-corrected yield of $D^0\to\pi^0\pi^0$ based on our
$\psi(3770)$ data set is $17521\pm500(stat.)\pm1559(syst.)$ events.


\clearpage

\subsection{$D^0\to\gamma\gamma$}

The analysis of $D^0\to\gamma\gamma$ starts by taking the most
and the $2^{nd}$ most energetic photon candidates in a given event,
where the photon selection criteria are the same as the ones for
the {\it good photons} previously described except we restrict these
two photons to be reconstructed only within the barrel section of
the EMC in order to suppress contaminations from continuum (i.e.,
$e^+e^-\to\gamma^*\to q\bar{q}\to\mbox{light hadrons}$, where $q=u,d,s$),
including doubly radiative Bhabha events.
Even though these photons are Doppler broadened in the lab frame,
the reconstructed photon energies, thus including the detector
resolutions, are mostly found to be at least $700$~MeV.

\begin{figure}[h]
\centering
\includegraphics[height=3.1in,trim=0mm 0mm 0mm 33mm,clip]{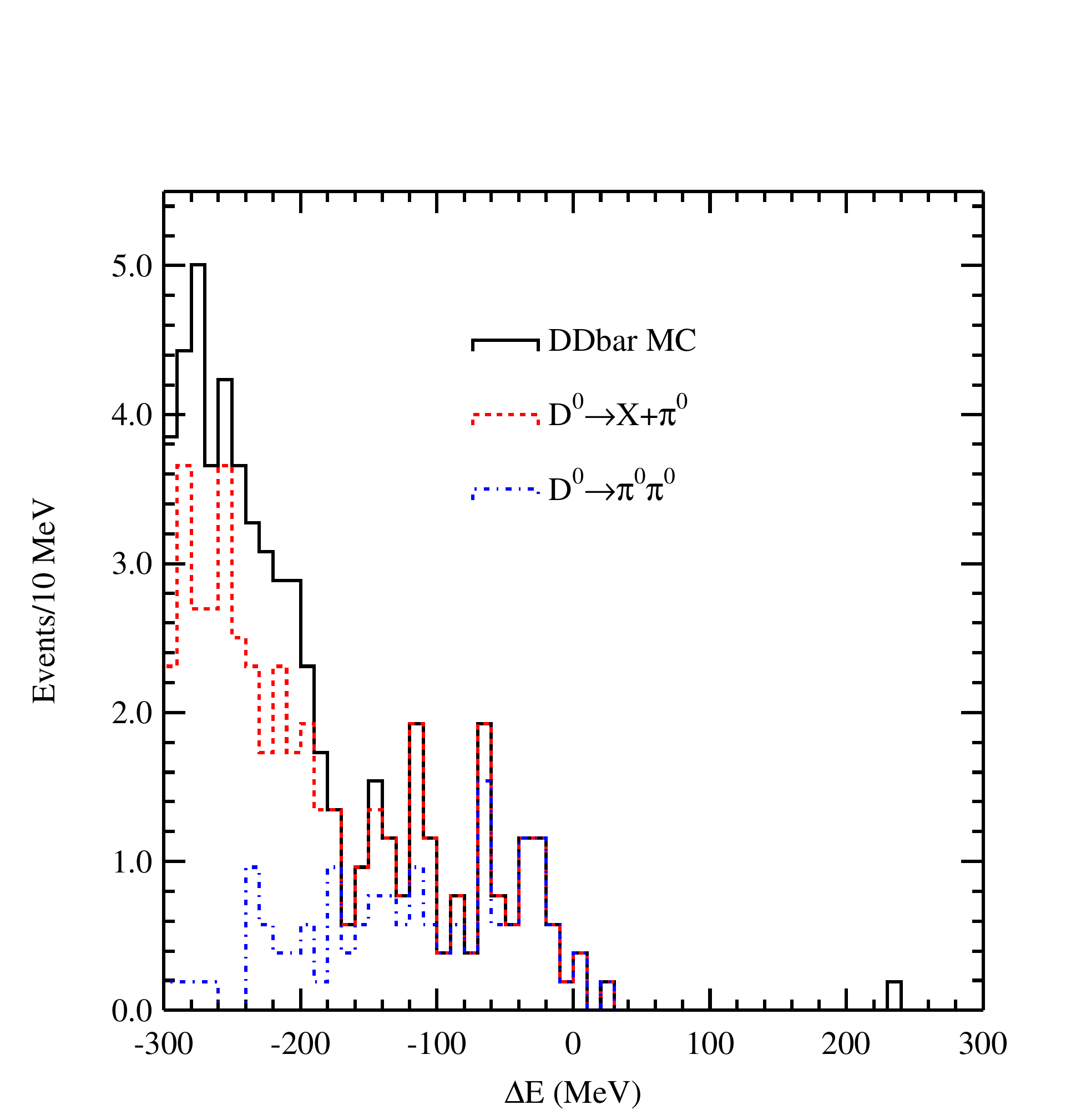}
\caption{$D^0\to\gamma\gamma$ candidate $\Delta E$ distribution based
 on MC sample
in which only generic decays of
$D\bar{D}$ are present, and $D^0\to\gamma\gamma$ is set to zero.
Notice that almost the entire background comes from two-body decays,
$D^0\to X+\pi^0$, denoted by the red-dashed histogram.
In the signal region ($|\Delta E| < 150$~MeV corresponds to $\sim\pm3\sigma$),
the background is dominated by events from $D^0\to\pi^0\pi^0$,
represented by the blue-dotted histogram.}
\label{fig:gamgam_background_pi0pi0}
\end{figure}

There are two major backgrounds in this analysis, one from the continuum processes
as mentioned above and the other from $D^0\to\pi^0\pi^0$. The later
poses an irreducible background when the two $\pi^0$s decay into
four photons in which two of them carry away most of the initial $D^0$
momentum.
Figure~\ref{fig:gamgam_background_pi0pi0} shows the $\Delta E$
distribution based on our MC sample in which only generic decays of
$D\bar{D}$ are present, and $D^0\to\gamma\gamma$ is set to zero.
It can be seen that the background in the signal region 
($|\Delta E| < 150$~MeV which corresponds to $\sim\pm3\sigma$) is dominated
by the events from $D^0\to\pi^0\pi^0$, represented by the blue-dotted
histogram in the figure.

To further suppress these backgrounds, our signal selection criteria
are optimized based on MC samples to maximize the signal sensitivity.
Each of the signal photon candidates must have a lateral shower
profile that is consistent with that of an isolated electromagnetic shower.
We suppress photons that look like coming from
$\pi^0\to\gamma\gamma$ by rejecting a candidate that forms
$\cos{\theta_{\gamma\gamma}}\ge0.6$  where $\theta_{\gamma\gamma}$ is an
opening angle between a signal photon candidate and any other photon
when the given pair form $110<M_{\gamma\gamma}<150$~MeV$/c^2$.
Requiring each of the signal photon candidates to be at least
$20^{\circ}$ away from any reconstructed charged track effectively suppresses
the doubly radiative Bhabha events. In addition, we demand all
selected events to satisfy $E_{EMC}/p<0.8$ or $E_{EMC}/p>1.05$, where
 $p$ is the momentum of the fastest reconstructed charged track in a
 given event and $E_{EMC}$ is the corresponding deposited energy in
 the EMC.  Electrons and positons tend to give $E_{EMC}/p\sim1$.
To suppress contaminations from the rest of the continuum processes,
we require there be at least one charged kaon reconstructed in a given event.
In the end, we extract our signals from a $\Delta E$ distribution
while requiring  $1860<M_{bc}<1870$~MeV$/c^2$ which gives
an overall reconstruction efficiency of $(12.02\pm0.1)\%$.

\begin{figure}[htb]
\centering
\includegraphics[height=4.0in,trim=0mm 0mm 0mm 33mm,clip]{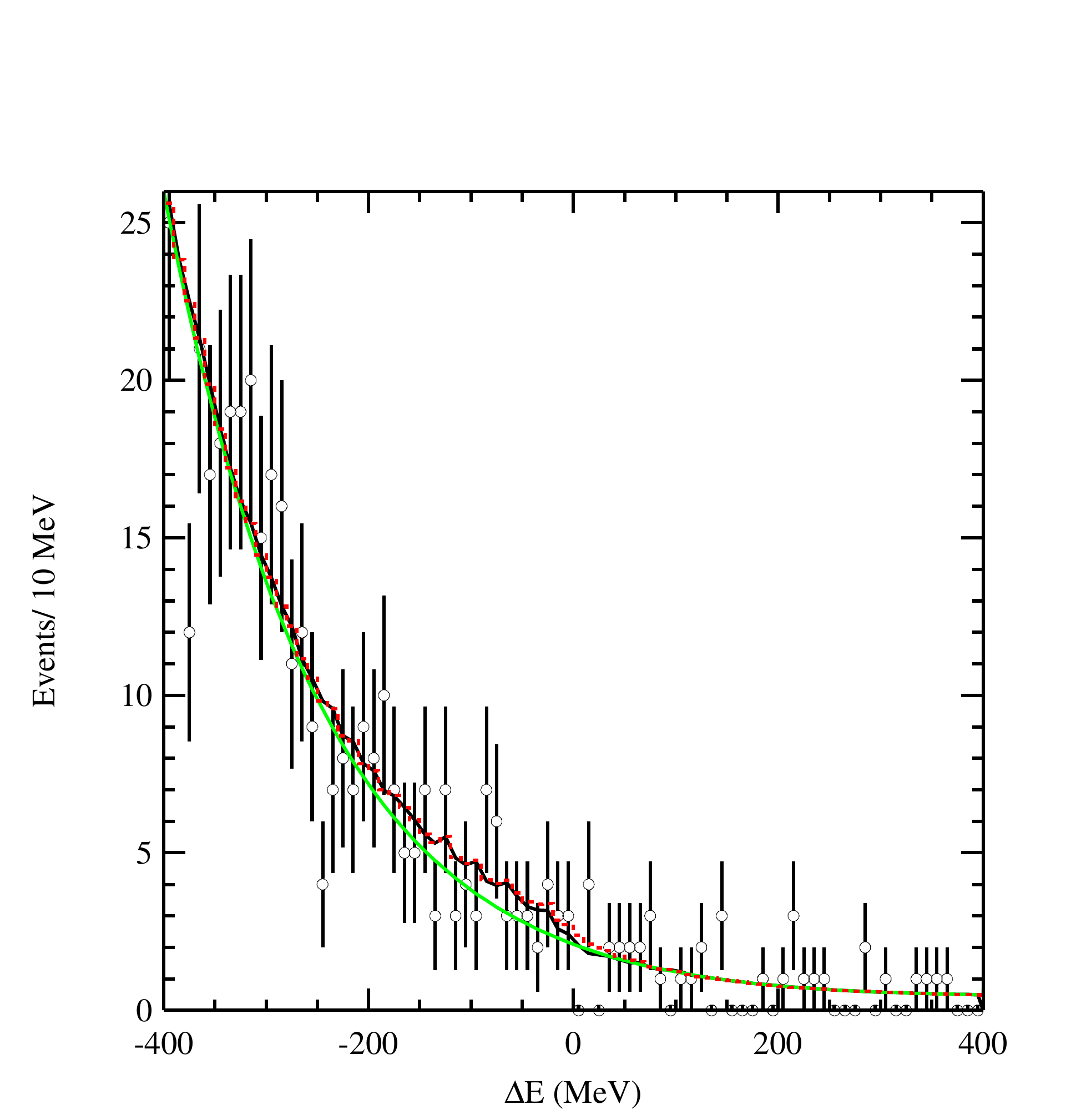}
\caption{A fit to the $D^0\to\gamma\gamma$ candidate $\Delta E$
  distribution based on the $\psi(3770)$ data sample.
Black points are data, solid
black curve is the overall fitted curve (signal plus backgrounds),
the red-dashed curve is the fitted total background, and the green
curve is the exponential and linear polynomials.}
\label{fig:gamgam_dtfit}
\end{figure}

Figure~\ref{fig:gamgam_dtfit} shows the result of a
 maximum-likelihood fit to the $\Delta E$
distribution based on the $\psi(3770)$ data set.
In the fit, the signal shape is fixed by the corresponding MC shape.
The background shape consists of three parts; MC-based shape to represent
the contamination from $D^0\to\pi^0\pi^0$ whose size is also fixed
based on our own observation; a $1^{st}$ order polynominal that covers
the contamination from 
Bhabha events which appear smoothly over the entire $\Delta E$
spectrum; a $1^{st}$ order exponential polynominal, corresponding to
the rest of the backgrounds.
Black points are data, the black-solid
curve is the overall fitted curve (signal plus backgrounds),
the red-dashed curve is the fitted total backgrounds, the green
curve is a sum of the exponential and linear polynomials.
The fit gives $\chi^2$ of 63.7 for 80 data points
(minus the 4 floated parameters) which yields
$-2.9\pm7.1$ signal events.
This translates into  an upper limit of $11$ events at $90\%$ confidence
level (C.L.) based on the Bayesian method.

Table~\ref{tab:gamgammsystdata} shows a summary of our estimation of the
systematic uncertainties on the ratio, $\br(D^0\to\gamma\gamma)/\br(D^0\to\pi^0\pi^0)$.
The uncertainties we consider include: photon reconstruction
efficiencies; the event-wise selections that require there be at least
one charged Kaon reconstructed and the cut on $E_{EMC}/p$; the
requirement on $M_{bc}$; the detector resolution of the signal shape;
the assumed background shape, fit range. In this table,
sources of uncertainties, that are estimated based on the data
and that are determined to be no more than
half of the measured statistical uncertainty, are listed as ``negligible''.

\begin{table}[htbp]
\caption{Summary of the estimated systematic uncertainties on
$\br(D^0\to\gamma\gamma)/\br(D^0\to\pi^0\pi^0)$.}
\label{tab:gamgammsystdata}
\begin{center}
\begin{tabular}{c| c }
Sources & rel. errors ($\%$)\\
\hline
Photon recon.    & $5.0$ \\
Event-wise cut   & $4.9$ \\
Cut on $M_{bc}$   & $2.5$ \\
Signal resolution & negligible \\
Background shape  & negligible \\
Fit range         & negligible \\
MC stat.          & 0.4 \\
Stat. from $D^0\to\pi^0\pi^0$ & 2.9 \\
Syst. from $D^0\to\pi^0\pi^0$ & 8.9 \\
\hline
Total  & 12.0 \\
\hline
\end{tabular}
\end{center}
\end{table}

Including the estimated total systematic uncertainty, we arrive at
$\br(D^0\to\gamma\gamma)/\br(D^0\to\pi^0\pi^0) < 5.8\times10^{-3}$ at
$90\%$ C.L.
\clearpage
\section{Conclusions and a future prospect}

Based on the ${\sim}2.9$ fb$^{-1}$ data set taken at $\sqrt{s}=3.773$~GeV,
we search for $D^0\to\gamma\gamma$.
We find no significant signal and set our preliminary upper limit on
$\br(D^0\to\gamma\gamma)/\br(D^0\to\pi^0\pi^0)<5.8\times10^{-3}$ at
$90\%$ C.L. With the known value of
$\br(D^0\to\pi^0\pi^0)$~\cite{pdg}, this corresponds to
$\br(D^0\to\gamma\gamma)<4.7\times10^{-6}$.

While we are waiting for BESIII to take more data at
$\sqrt{s}=3.773$~GeV, there is an alternate analysis approach that is
unique to our data sample. The produced $\psi(3770)$ in our sample
decays into a pair of $D^0\bar{D^0}$. Reconstructing one of the
$D^{0}$ mesons with known exclusive modes while searching for
$D^0\to\gamma\gamma$ in the other $D^0$ decay would yield an almost
background-free environment, except for the irreducible contamination from
$D^0\to\pi^0\pi^0$ for which we have control. Such a study is also
currently under way.


\Acknowledgements
I thank J. L. Rosner for useful comments on the manuscript.


\begin{thebibliography}{99}

\bibitem{gim1970} S. L. Glashow, J. Iliopoulos, and L. Maiani,
  Phys.\ Rev.\ D {\bf 2}, 1285 (1970).

\bibitem{greub1996} C. Greub, T. Hurth, M. Misiak, and D. Wyler,
  Phys.\ Lett.\ B {\bf 382}, 415 (1996).

\bibitem{fajfer2001} S. Fajfer, P. Singer, and J. Zupan, Phys.\ Rev.\ D {\bf 64}, 074008 (2001).

\bibitem{burdman2002} G. Burdman, E. Golowich, J. L. Hewett, and
  S. Pakvasa, Phys.\ Rev.\ D {\bf 66}, 014009 (2002).

\bibitem{Prelovsek2001} S. Prelovsek and D.Wyler, Phys.\ Lett.\ B {\bf
    500}, 304 (2001).

\bibitem{bigi2010} A. Paul, I. I. Bigi, and S. Recksiegel,
  Phys.\ Rev.\ D {\bf 82}, 094006 (2010).

\bibitem{cleo} T. E. Coan {\it et al.} (CLEO Collaboration),
  Phys.\ Rev.\ Lett.\ {\bf 90}, 101801 (2003).

\bibitem{babar} J. P. Lees {\it et al.} (BaBar Collaboration), Phys.\ Rev.\ D {\bf 85}, 091107 (R) (2012).

\bibitem{besiii} M. Ablikim {\it et al.} (BES Collaboration), Nucl.\ Instrum.\ Meth.\ A {\bf 614}, 345 (2010).

\bibitem{pdg} K. Nakamura {\it et al.} (Particle Data Group), Journal of
  Physics G{\bf 37}, 075021 (2010) and 2011 partial update for the 2012
  edition.

\end{thebibliography}
\end{document}